\documentclass{IEEEtran}

\usepackage{cite,graphicx,amsmath,amssymb,hyperref,url,color,multirow}
\usepackage{cases}
\usepackage{ulem}
\usepackage{epstopdf}
\usepackage{listings}
\usepackage{tikz}

\ifCLASSOPTIONcompsoc
\usepackage[tight,normalsize,sf,SF]{subfigure}
\else
\usepackage[tight,footnotesize]{subfigure}
\fi

\definecolor{awesome}{rgb}{1.0, 0.6, 0.0}

\newcommand\copyrighttext{
{\copyright} 2018 IEEE. Personal use of this material is permitted. Permission from IEEE must be obtained for all other uses, in any current or future media, including reprinting/republishing this material for advertising or promotional purposes, creating new collective works, for resale or redistribution to servers or lists, or reuse of any copyrighted component of this work in other works. \textbf{This paper will appear in the August 2018 issues of IEEE Wireless Communications Magazine - Network Testing and Analytics}
}
\newcommand\copyrightnotice{%
	\begin{tikzpicture}[remember picture,overlay]
	\node[anchor=south,yshift=10pt] at (current page.center) {\fbox{\parbox{\dimexpr\textwidth-\fboxsep-\fboxrule\relax}{\copyrighttext}}};
	\end{tikzpicture}%
}
\begin{document}

\title{COINS: ContinuOus IntegratioN in wirelesS technology development}

\author{
\IEEEauthorblockN{Matev\v z Vu\v cnik \IEEEauthorrefmark{1}\IEEEauthorrefmark{2}, Toma\v z \v Solc \IEEEauthorrefmark{1}\IEEEauthorrefmark{2}, Urban Gregorc \IEEEauthorrefmark{3}, Andrej Hrovat \IEEEauthorrefmark{1}\IEEEauthorrefmark{2}, Klemen Bregar \IEEEauthorrefmark{1}\IEEEauthorrefmark{2}, Miha Smolnikar \IEEEauthorrefmark{1}, Mihael Mohor\v ci\v c \IEEEauthorrefmark{1}\IEEEauthorrefmark{2} and Carolina Fortuna \IEEEauthorrefmark{1}}\\
\IEEEauthorblockA{\IEEEauthorrefmark{1}Jo\v zef Stefan Institute, Ljubljana, Slovenia\\
\IEEEauthorrefmark{2}Jo\v zef Stefan International Postgraduate School, Ljubljana, Slovenia\\
\IEEEauthorrefmark{3}ComSensus Ltd., Ljubljana, Slovenia\\
\{matevz.vucnik, tomaz.solc, andrej.hrovat, klemen.bregar, miha.smolnikar, miha.mohorcic, carolina.fortuna\}@ijs.si, urban.gregorc@comsensus.eu}}

\copyrightnotice

\maketitle

\begin{abstract}

	Network testing plays an important role in the iterative process of developing new communication protocols and algorithms.
	
	However, test environments have to keep up with the evolution of technology and require continuous update and redesign.
	In this paper, we propose COINS, a framework that can be used by wireless technology developers to enable continuous integration (CI) practices in their testbed infrastructure. As a proof-of-concept, we provide a reference architecture and implementation of COINS for controlled testing of multi-technology 5G Machine Type Communication (MTC) networks. The implementation upgrades an existing wireless experimentation testbed with new software and hardware functionalities. It blends web service technology and operating system virtualization technologies with emerging Internet of Things technologies enabling CI for wireless networks. Moreover, we also extend an existing qualitative methodology for comparing similar frameworks and identify and discuss open challenges for wider use of CI practices in wireless technology development.
\end{abstract}

%
\IEEEpeerreviewmaketitle

\begin{IEEEkeywords}
	continuous integration, wireless networks, software development, testbed, machine type communication
\end{IEEEkeywords}

\section{Introduction}

\begin{figure}[htb]
	\centering
	\includegraphics[width=\columnwidth]{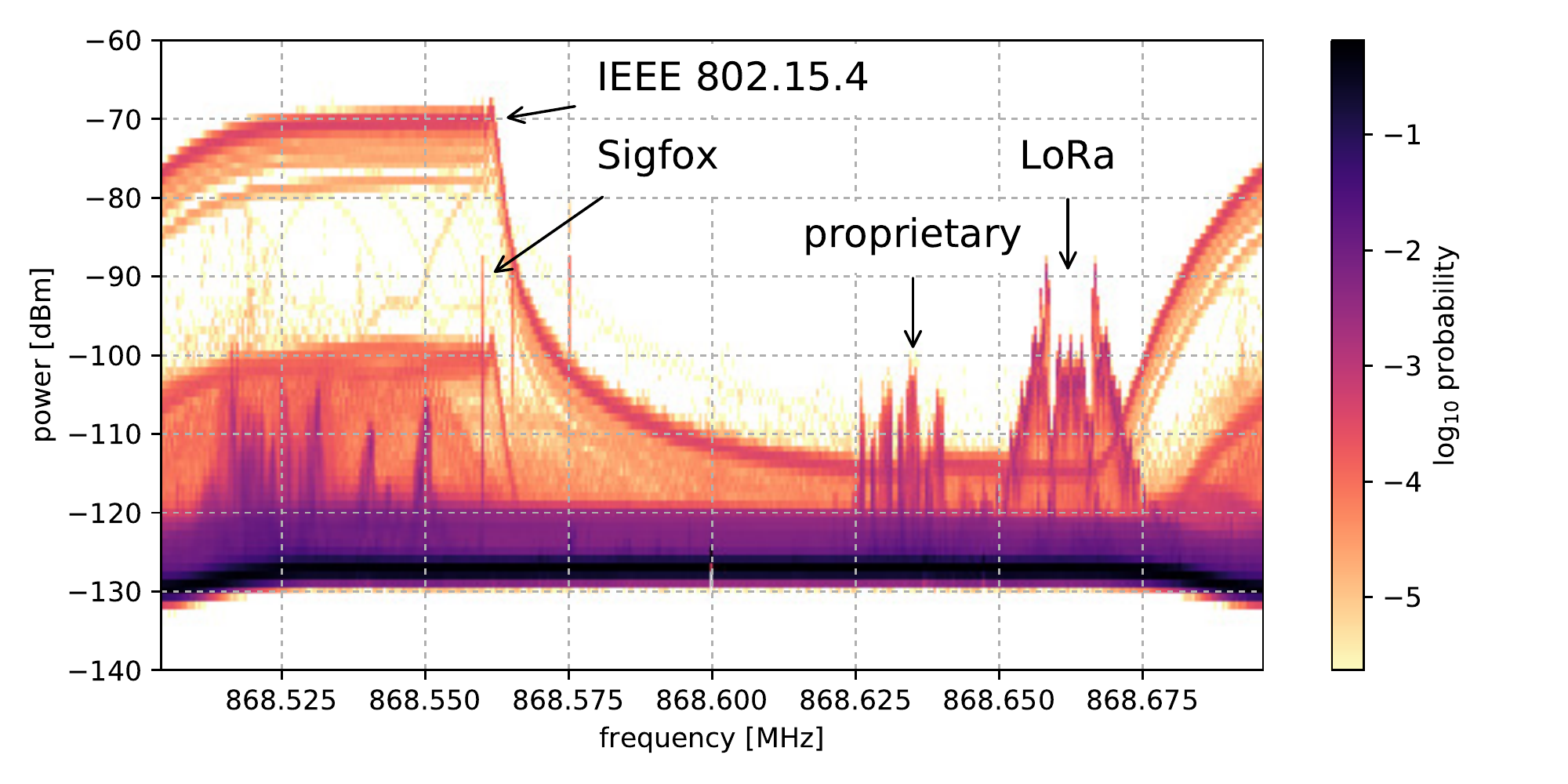}
	\caption{A histogram of power spectral density samples over a 200 kHz wide band in the unlicensed European 868 MHz SRD band in Ljubljana, Slovenia.}
	\label{fig:spectrum}
\end{figure}

The next generation of mobile networks will have to support industrial and traffic safety applications, enhanced multimedia applications on smartphones and a plethora of connected meters, sensors and other devices that will comprise living and working environments in the future. In order to meet the requirements of all these applications, mobile networks will have to accommodate both human-type communications (HTC) and machine-type communications (MTC) \cite{condoluci2015toward}. MTC capabilities of existing predominantly HTC-oriented cellular technologies will be enhanced through the inclusion of non-cellular short-range and low-power wide area technologies under the 5G umbrella in the form of capillary networks. This will result in a multi-technology, high-interference radio environment \cite{DePoorter2017} such as can be found in some frequency bands in dense urban centers. In Figure \ref{fig:spectrum}, spectrum measurements in the shared 200 kHz unlicensed sub-1 GHz band over the course of 24 hours show signals from IEEE 802.15.4-based, LoRa and Sigfox networks as well as a number of unidentifiable proprietary technologies.

Given the increasing complexity of wireless technologies, evaluation and testing of new devices, protocols or solutions in realistic environments is desirable but tedious and expensive. Although complicated and expensive to build and maintain, distributed and heterogeneous wireless testbeds offer the most realistic experimental conditions \cite{5723816}. While the cost of adding additional hardware functionality to support emerging technologies is difficult to overcome, the cost of developing software for the increased variety and number of wireless devices can be significantly reduced using continuous integration (CI) principles and tools \cite{meyer2014continuous}. In particular, open source software projects have benefited significantly from adopting CI principles that significantly reduce the workload of the core maintainers. Contributed code can be automatically tested for syntax, style and errors using sophisticated toolchains that automate previously manual development practices. Studies have shown that the CI methodology significantly increases productivity and code quality \cite{vasilescu2015quality}.

A recently proposed software-centric systematic testing methodology for mobile networks \cite{Pinola13} incorporates unit testing, integration testing and system testing, which are central to software development and represent an important part of the CI process. Testbed support enabling such a methodology can be economically implemented using modern tools, such as software containers, that can be automatically deployed to test new code.

The main contribution of this paper is COINS, a framework for ContinuOus IntegratioN in wirelesS technology development that can be used by developers to enable conventional CI practices on wireless infrastructure. To illustrate and facilitate discussion on the framework capabilities, we also provide a reference architecture and implementation. In particular, the existing LOG-a-TEC testbed \cite{vsolc2015low, fortuna2017software} for controlled testing of emerging multi-technology 5G capillary MTC networks was upgraded to enable CI. LOG-a-TEC is a combined outdoor and indoor heterogeneous wireless testbed for experimentation with sensor networks and machine-type communication\footnote{LOG-a-TEC, \url{http://log-a-tec.eu/}}.

Two further contributions include (i) extending an existing qualitative methodology \cite{Buchert} for comparing similar frameworks; and (ii) identification of and discussion on open challenges for wider adoption of CI practices in the development of wireless technologies.

The paper is structured as follows. Section \ref{sec:proc} identifies the main functionalities of a wireless experimentation testbed required by a software developer and explains how the CI process can be utilized in handling them. Section \ref{sec:architecture} details the reference architecture and implementation of COINS. Section \ref{sec:deployment} discusses the deployment of the new wireless experimentation technology as an extension of the existing LOG-a-TEC testbed and lists examples of supported experiments, while Section \ref{sec:related} compares COINS with related work. Section \ref{sec:challenges} identifies and discusses open challenges. Finally, Section \ref{sec:conclusions} summarizes the paper.



\section{Continuous integration process on wireless testbeds}
\label{sec:proc}
There are four main functionalities that wireless network developers and testers need from a testbed environment:

\begin{itemize}
	\item Simple, low-effort integration of software and hardware components that require testing; the priority is to minimize the overhead that testbed usage introduces into the technology development cycle.
	\item Controllable generation of traffic in the network under test, such as on-demand packet transmission according to custom patterns and distributions, to test new network protocols under realistic traffic conditions.
	\item Controlled emulation of a radio operating environment using various technologies in order to identify potential coexistence issues between new and legacy technologies as early in the development cycle as possible.
	\item Monitoring and analysis of network and radio spectrum metrics.
\end{itemize}

The introduction of the CI process into the wireless technology development proposed in this paper significantly lowers the barriers to the first and last of these functionalities while relying on the other two to successfully carry out the CI tests. According to existing CI practices \cite{Duvall}, the proposed process should properly interface three main entities depicted in Figure \ref{fig:testlab-architecture}a:
\begin{enumerate}
	\item \textit{Repository} that holds the code, configuration files, tests and documentation, i.e., everything needed for the automated build. Generally, the repository is implemented using a version control system such as Git or Mercurial.
	\item \textit{Automation system} that performs all the instantiations and builds and transfers software components. Conceptually, it does for CI what the production line does for a factory. Its implementation typically includes a large set of tools and components. In modern software development, the most efficient approach is to use web and containerization technologies.
	\item \textit{Testbed device} is used to run the tests and experiments. It can be off-the-shelf equipment, a custom built device or a hybrid. The first approach is more suitable for improving existing technologies, while the second is for developing new ones. For the sake of generality, we assume in the following that a testbed device comprises separate functional blocks represented by the infrastructure node for device management and the target node for test execution.
\end{enumerate}

As depicted in Figure \ref{fig:testlab-architecture}a, the CI process begins in step 1 when a new testbed device automatically registers itself in the automation system and becomes available for testing. In step 2, the developer sends new code (pushes a commit) to the repository, initiating step 3, which consists of triggering the automation system. In step 4, the automation system pulls the changes from the repository and executes the deployment scripts specifying the testbed devices required by the test case. Step 5 deploys the changes from the repository to the testbed device and instructs it to start the build process. In step 6, the testbed device initiates the build process on the infrastructure node. The build system is contained inside a container; therefore, the infrastructure part of the testbed device stays unaffected by any build failures or run-away processes. If the building process succeeds, the infrastructure node in step 7 flashes the target node with freshly built firmware and in step 8 executes the test process. In step 9, the infrastructure node obtains the results from the target node and examines them. In step 10, the automation system updates the repository with the result and communicates the result to the developer, typically via email.

\section{COINS architecture and implementation}
\label{sec:architecture}

\begin{figure*}[htb]
	\centering
	\includegraphics[width=.95\textwidth]{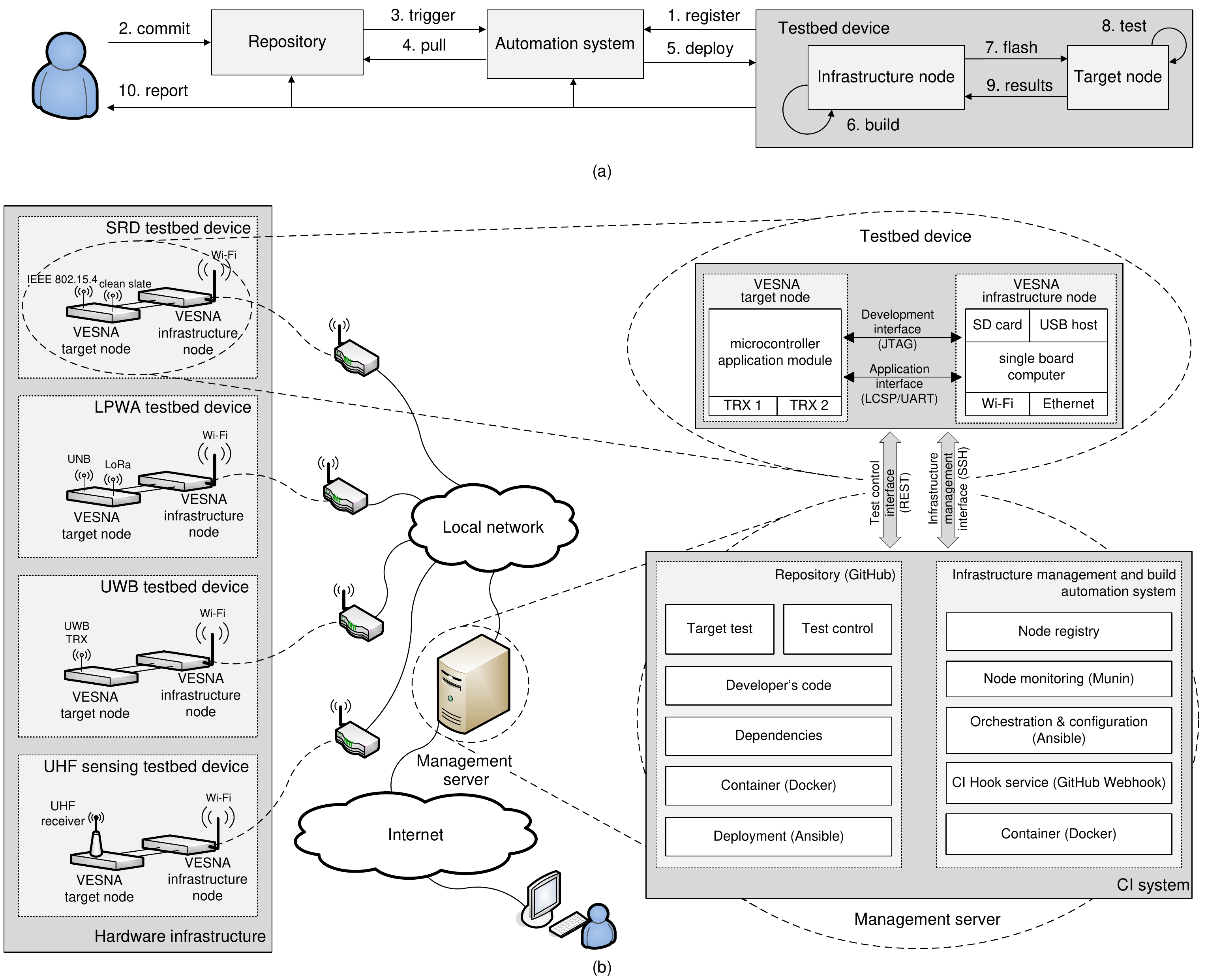}
	\caption{COINS framework: a) CI process; b) Architecture of the reference implementation.}
	\label{fig:testlab-architecture}
\end{figure*}

The architecture of COINS is illustrated in Figure \ref{fig:testlab-architecture}b and further explained in this section. It consists of the three main architectural components identified and described in Section \ref{sec:proc}, which are mapped on the elements of the existing LOG-a-TEC infrastructure \cite{vsolc2015low, fortuna2017software}. The COINS architecture is implementation independent with well defined modules which interact through specified interfaces for which we provide a reference implementation although another implementation using a set of different tools is certainly possible.

For the reference implementation of COINS\footnote{COINS, \url{ https://github.com/sensorlab/SensorManagementSystem}}, we chose widely used open source tools. In this way, COINS benefits from and contributes to the work of established communities and is more likely to be adopted. In particular, the container system is based on Docker\footnote{Docker, \url{https://www.docker.com}}, which we use to package software for distribution to nodes in the testbed. The CI hook service adds support for the GitHub WebHook\footnote{GitHub, \url{https://developer.github.com/webhooks}} client. We use Ansible\footnote{Ansible, \url{https://www.ansible.com}} to describe and automate tasks. Ansible is a popular automation engine for configuration management and software deployment. The networked resource monitoring tool Munin\footnote{Munin, \url{http://munin-monitoring.org}} serves for testbed device monitoring. We also implemented a custom node registry system and released it as free software under the AGPL license.

\subsection{Testbed devices}
\label{sec:hw}
The LOG-a-TEC testbed devices can use several wireless technologies, as depicted on the left side of Figure \ref{fig:testlab-architecture}b. Architecturally, we consider hybrid testbed devices with two separate functional blocks represented by the infrastructure node and the target node. The main reason for the separation of the testbed device into two functional blocks is to have a generic infrastructure node that can be combined with various target nodes to support a range of heterogeneous experiments with a unified way of management and reconfiguration. Another reason is to achieve hardware separation between robust and stable testbed management functionality on the infrastructure node and the experimental code, which is prone to errors and crashes on the target node. The infrastructure node can usually also support containerization, while the target node is often based on an embedded system with limited hardware resources.

In our implementation, the infrastructure node is a custom-designed single board computer based on the BeagleCore module running the Debian GNU/Linux operating system. It features wired Ethernet and Wi-Fi for infrastructure connectivity, an SD card slot for storage expansion, VESNA and USB interfaces for modular extensibility and a development/debug process. The target node is a custom VESNA (VErsatile platform for Sensor Network Applications) device with an ARM Cortex-M3 microcontroller application module and dedicated experimentation transceivers. VESNA can run a dedicated OS (e.g., Contiki) or custom firmware. The integration of the two is done via custom hardware design and the adoption of selected interfaces and protocols, thus going beyond similar approaches such as \cite{7919566}. Both nodes are interconnected via application and development interfaces, providing the exchange of application data as well as remote low-level application debugging. The reliable development interface is based on JTAG and the application interface on protocols such as LCSP (Light-weight Client Server Protocol; an HTTP-inspired protocol \cite{vsolc2015low}) running on top of a serial interface.

\subsection{Repository}
\label{sec:repo}
The repository support was added to the system to abstract the underlying complexity and enable the CI process described in Section \ref{sec:proc}. It includes (i) the target test that has to be executed on the testbed device; (ii) the test control that runs on the management server and communicates with the testbed device; (iii) the developer's code that has to be tested; (iv) dependencies such as software libraries; (v) the container file, which describes how to package the target test, test control, dependencies and developer's code into a container; and (vi) the deployment configuration, which specifies where and how to build and run the container.

In our implementation, the target test and test control are realized using C and Python, but in general, the proposed framework is programming language independent. The container support is implemented through Docker, while the deployment is automated by Ansible. The repository that contains everything needed for the completely automated build is realized through a GitHub repository. Examples of more advanced tests and experiments are summarized in Section \ref{sec:deployment}.

\subsection{Infrastructure management and build automation system}
\label{sec:management}

The infrastructure management and build automation system is a complex setup composed of self-sufficient systems packaged in a Docker container following the micro-service architecture approach. It contains (i) the node registry, which has a list with details about all testbed devices; (ii) the node monitoring service, which is in charge of monitoring the activity and health of the testbed devices; (iii) the orchestration and configuration, which performs regular maintenance and reconfiguration of the testbed; (iv) the CI hook service, which waits for signals of changes in the repository and initiates the build process; and (v) the container file, which describes how to package all the previous components in a container.

There are several periodic background jobs running on the management system that perform house-keeping tasks:
\begin{itemize}
	\item{building the Ansible host file from the node registry database,}
	\item{updating the information on the availability of individual nodes using the Ansible ping command,}
	\item{stand-alone testbed device monitoring based on Munin, and}
	\item{CI Hook service triggering the build process.}
\end{itemize}

Externally, the management server exposes a user-facing web interface and an HTTP REST API \cite{fortuna2017software}. The infrastructure management and build automation system allows browsing through clusters of individual devices and nodes in the database, positioning devices on a map, visualizing reports, monitoring device activity, etc. Thus, it gives an instant overview of the state of the testbed at any time. Warnings can be defined to be triggered when specific monitored variables exceed a given threshold, such as running out of system memory or storage space. In such cases, the system sends an email to the system administrator. Depicting the testbed devices on a map is particularly useful for the selection of devices for different wireless test cases. Programmatic access to the management system is performed through the REST API; i.e., the nodes use the REST API to perform automatic registration and update changes in the configuration.

\section{Testbed deployment and supported functionality}
\label{sec:deployment}

\begin{figure}[htb]
	\centering
	\includegraphics[width=.95\linewidth]{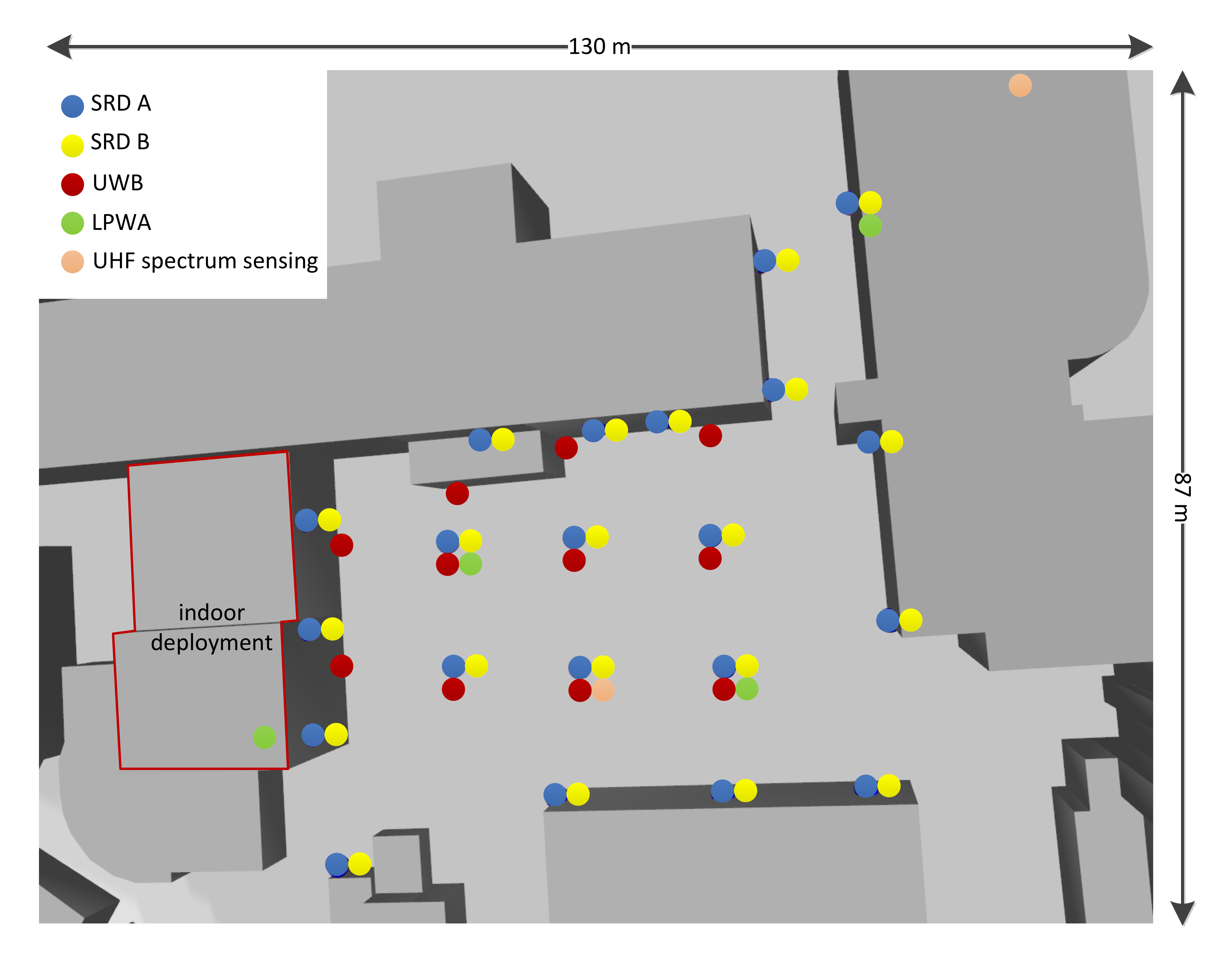}
	\caption{Locations of available testbed devices in outdoor environment.}
	\label{fig:deployment}
\end{figure}

In order to enable testing and experimentation with the COINS framework in a dense and heterogeneous 5G capillary radio environment, we upgraded the combined outdoor/indoor LOG-a-TEC testbed at the JSI campus with the necessary tools and services described in Section \ref{sec:architecture}. Table \ref{tab:deployment} summarizes the number and type of testbed devices in an indoor and outdoor environment, whereas the layout of devices in the outdoor environment is depicted in Figure \ref{fig:deployment}.

\begin{table}[htb]
	\renewcommand{\arraystretch}{1.1}
	\caption{Configuration and features of testbed devices.}
	\label{tab:deployment}
	\centering
	\begin{tabular}{llll}
		\bfseries Node type & \bfseries Radio & \bfseries Frequency & \bfseries No. of dev. \\
		[1ex] \hline \\ [-1ex]
		\multirow{2}*{Target node SRD A}	& Atmel AT86RF212	& 868 MHz	& \multirow{2}*{21 outdoor} \\
					& TI CC2500		& 2.4 GHz	& \\
		[1ex] \hline \\ [-1ex]
		\multirow{2}*{Target node SRD B}	& TI CC1101		& 868 MHz	& \multirow{2}*{21 outdoor} \\
					& Atmel AT86RF231	& 2.4 GHz	& \\
		[1ex] \hline \\ [-1ex]
		\multirow{2}*{Target node LPWA}	& \multirow{2}*{Semtech SX1272}	& \multirow{2}*{868 MHz}	& 3 outdoor \\
		& & & 1 indoor \\
		[1ex] \hline \\ [-1ex]
		\multirow{2}*{Target node UWB}	& \multirow{2}*{DecaWave DW1000}	& 3.5 - 6.5 GHz		& 11 outdoor \\
					&					& (6 channels)		& 20 indoor \\
		[1ex] \hline \\ [-1ex]
		Target node		& \multirow{2}*{NXP TDA18219HN}			& \multirow{2}*{470 - 866 MHz}		& \multirow{2}*{2 outdoor} \\
		UHF spec. sens. \\
		[1ex] \hline \\ [-1ex]
		Infrastructure		& \multirow{2}*{TI WL1837}		& \multirow{2}*{5 GHz}	& 58 outdoor \\
		node			&					&			& 21 indoor \\
	\end{tabular}
\end{table}

The outdoor part of the wireless testbed is located in and around a park area of 55 m by 60 m. Testbed devices are placed on light poles 3.5 m above the ground and on the surrounding buildings at heights from 2.0 to 9.3 m. To enable experiments in indoor as well as in mixed indoor/outdoor scenarios, the testbed is extended in an indoor environment with additional 20 UWB devices and one LPWA device. These are deployed on the second and third floors of the 28.4 m by 16.6 m building. In addition, several portable devices of all types described in Section \ref{sec:hw} are at disposal for analyzing the use cases incorporating device mobility or further network densification.

The diversity of types, the relatively high number and density of testbed devices, and the possibility of hosting multiple radio technologies in most devices to transmit/receive/sense at the same time at the same location enables experimentation scenarios in a very dense and heterogeneous wireless environment. The various types of experiments include: (i) investigation of assisted and non-assisted indoor/outdoor radio localization algorithms; (ii) spectrum sensing for dynamic spectrum access and cognitive radio networking; (iii) link and/or network performance evaluation in a controlled interference and/or heterogeneous radio environment; (iv) design and evaluation of new custom-developed wireless protocols and their benchmarking against standardized protocols; and (v) performance evaluation and controlled piloting of prototype IoT devices.

In order to support such diverse testing capabilities, the repository comprises a set of testbed libraries and tools. They enable: out of the box localization; Line-of-Sight/non-Line-of-Sight link classification; link quality estimation (LQE), classification and prediction; custom protocol stack development using modular components; spectrum sensing by energy detection, cyclostationary detection and eigenvalue- and covariance-based methods; signal generation including wireless microphone profiles; and game theoretic power allocation. These software components and modules can be incorporated in custom developed algorithms or adapted to the particular use case and subsequently used as a testing benchmark. They can also be used for setting the required testing environment. In particular, for evaluation and comparison of spectrum sensing algorithms and other wireless technologies or algorithms, the testbed enables generating controlled interference. Furthermore, localization and distributed spectrum sensing can be used to build geolocation spectrum occupancy databases, which in turn can be used in investigation of primary/secondary spectrum access regimes.

A minimum test case for a wireless communication system includes software for three entities: the transmitter, the receiver and the test controller. The first two will run on the target node within the testbed device, while the third runs on the management server. The test controller instructs one testbed device to go in receive mode and another testbed device to transmit the test sequence. When the first testbed device receives the test sequence, it reports the results back to the test controller, which asserts that the received sequence matches the transmitted one. If it does, the test succeeds; otherwise, it fails i.e., \lstinline[basicstyle=\ttfamily, language=Python]{assertEqual(nodeTX.data, nodeRX.data)}. The described testing approach was implemented in our testbed for the purpose of LoRa firmware development. Especially in industry, there are situations where multiple developers work on the same code-base simultaneously. In such environments it can happen that a fix from one developer breaks a feature of another. By introducing automated testing on a real testbed, the integration problems get discovered and fixed early in the development process.

\section{Comparison with related work}
\label{sec:related}
While we did not find any other framework addressing wireless experimentation with CI support, we have identified three existing frameworks that are to some extent similar to COINS, i.e., the control and Management Framework (OMF) \cite{Rakotoarivelo}, the Network Implementation Testbed Laboratory (NITOS) \cite{7151032} and the Berlin Open Wireless Lab (BOWL) \cite{Fischer}. We performed a feature-wise comparison of these frameworks and summarized it in Table\,\ref{tab:compare}.

In the comparison, we considered an extensive list of features that can be used to compare the experimentation systems \cite{Buchert}. The list of features, however, lacks the CI-specific properties that our work focuses on; therefore, we extended the list with four core CI concepts \cite{Duvall,STAHL201448}:
\begin{itemize}
	\item Existence of a single source repository that contains everything needed for the completely automated build process.
	\item Support for completely automated build process executed on each commit.
	\item Self-tests included inside the repository, which run on each build.
	\item Fast build process so that each commit can be built and tested.
\end{itemize}

\begin{table}[htb]
	\centering
	\caption{Framework comparison.}
	\label{tab:compare}
	\begin{tabular}{lcccc}
		\bfseries \hfil Features & \bfseries OMF & \bfseries NITOS & \bfseries BOWL & \bfseries COINS \\
		\hline
		Description Language & OEDL & REST & UCI & Ansible/Docker \\
		Experiment Types & Any & Any & Wireless & Wireless \\
		Operation & Testbed & Testbed & Production/Testbed & Testbed \\
		Interoperability & \checkmark & \checkmark & & \checkmark \\
		Reproducibility & \checkmark & \checkmark & \checkmark & \checkmark \\
		Fault Tolerance & \checkmark & \checkmark & & \checkmark \\
		Debugging & \checkmark & \checkmark & \checkmark & \checkmark \\
		Monitoring & \checkmark & \checkmark & \checkmark & \checkmark \\
		Data Management & \checkmark & & & \checkmark \\
		Source Repository & & & & \checkmark \\
		Automated Build & & & & \checkmark \\
		Self Testing & & & & \checkmark \\
		Fast Build & & & & \checkmark \\
		\hline
	\end{tabular}
\end{table}

For each feature presented in Table\,\ref{tab:compare}, we explain how we addressed the problem compared to the other frameworks. The last four features are CI specific and are realized only in COINS.

\paragraph{Description Language}
The description language is the language used to describe how and where to build and run the experiment.
In OMF, the experiment is described using OMF Experiment Description Language. The NITOS testbed is abstracted through a REST API. The BOWL system is based on the OpenWRT Unified Configuration Interface (UCI) scripts.
In COINS we did not want to introduce a new language specifically for this purpose. Hence we used well-known tools that many developers already use in their development practices. In order to define what tests to run on which node and how to obtain test results we use the Ansible Playbook, a language for describing software deployment steps. For building our containers we use Dockerfile, a text format used by Docker to define container build procedures.

\paragraph{Experiment Types} The type of experiment describes the platform where the experiment is going to run. OMF and NITOS can support wired or wireless experiments, whereas BOWL is limited to WiFi-based wireless systems. COINS focuses on wireless systems and is suitable also for restricted capability devices. While OMF, NITOS and COINS are suitable for developing completely new as well as optimizing existing wireless solutions, BOWL is primarily suitable for optimizing existing solutions.

\paragraph{Operation} The operation refers to the ability of the testbed to also serve as a production system. OMF and NITOS are exclusively testbed frameworks, while BOWL is designed to work simultaneously as a testbed and production WiFi access network. In this respect, COINS is also exclusively a testbed framework.

\paragraph{Interoperability} Framework interoperability covers an infrastructure independent design, support for existing infrastructure services, resource discovery and software interoperability. OMF and NITOS address interoperability issues, while BOWL UCI configuration scripts are implementation specific. OMF can be extended, and NITOS specifies a testbed API. COINS addresses interoperability by offering a modular and implementation independent design.

\paragraph{Reproducibility} Reproducibility refers to the ability to automatically run the same experiments multiple times. OMF enables complete reproducibility, while NITOS offers limited reproducibility since it implements automated node selection based on availability. BOWL also addresses the problem of making experiments reproducible. COINS supports reproducibility of test cases by creating tags inside the repository and manually choosing the devices used in the test. While declaratively all frameworks address reproducibility, we argue that it may be severely affected in case of missing repository support on the framework level and challenged by the ability to control the conditions in the physical environment.

\paragraph{Fault Tolerance} Fault tolerance addresses recovery capabilities when the experiment does not execute as expected. OMF and NITOS frameworks are designed to be fault tolerant. In OMF, there is a Management Plane responsible for rebooting a resource and setting up particular disk images on PC-based resources. NITOS uses an automatic monitoring tool to diagnose a malfunction in one of the resources. BOWL is not fully fault tolerant as it also serves as a production WiFi network for actual users. COINS is inherently fault tolerant due to its architecture design having separated infrastructure and target node as well as separate development and application interfaces.

\paragraph{Debugging} The debugging process addresses the ability to identify problems and their causes during the experiment execution. All compared frameworks are designed to produce extensive log files during the build process. COINS also collects extensive debug log during the execution. Furthermore the debug log is automatically emailed to the developer in case the execution fails.

\paragraph{Monitoring} Monitoring is a feature that enables the developer or a maintainer to oversee the status of the system at any time. All compared frameworks have a system for monitoring. COINS uses the Ansible ping method, which works on top of SSH to identify which nodes are accessible. Furthermore, it integrates a standalone open-source monitoring tool Munin as part of the management system.

\paragraph{Data Management} Data management is responsible for the distribution and collection of data. OMF uses the Measurement Collection system based on a relational database. NITOS and BOWL do not prescribe any dedicated data collection mechanism on the framework level. COINS uses the same system for build data management as for debugging; i.e., each build produces the result files along with the debug files in the build repository.

\paragraph{Source Repository} The COINS framework is designed to keep everything needed for a completely automated build inside a single version-controlled source repository. This is in line with one of the fundamental CI concepts.

\paragraph{Automated Build} The automated build is another core CI concept. It ensures that a new code committed by the developer does not break the process, which is particularly important in CI. We showed that COINS defines the automated build system, which deploys the modified source code to the appropriate targets where it is built and tested.

\paragraph{Self Testing} Tests of each committed change are extremely important for the CI process. In COINS, all the tests are included inside the repository. The test process is executed automatically on each commit, and the result is communicated to the developer.

\paragraph{ Fast Build} The last core CI concept requires a fast build process so that the developer does not have to wait long for the test results. To achieve fast builds, COINS relies on the container technology and incremental changes delivery to the testbed device. Therefore, only the changes are rebuilt within the container, thus making the build process fast.

\section{Discussion on open Challenges}
\label{sec:challenges}

For wider adoption of CI practices in the development and testing of wireless technologies, we identified three groups of challenges. The first group revolves around the constraints posed by the physical environment; the second one is concerned with process automation; and the third group is related to the lifecycle management of the wireless product.

\subsection{Challenges related to the physical environment}
A CI system for wireless networks comprised of distributed hardware has to take into account that there may be unpredictable radio interference on the physical transmission medium or failures in hardware, and it must be ensured that the test reports to the developer are immune to them. In other words, from the test reports, it should be clear if the test failed because of software issues or because of conditions in the physical environment.

\paragraph{Reproducibility} A possible approach for ensuring reproducibility of the test results in the unpredictable wireless environment is for the CI framework to use spectrum sensing to identify a suitable test channel and thus avoid interference. Additionally, using controlled interference in test cases can sometimes prove useful to verify the interference specific performance of a particular wireless algorithm.

\paragraph{Reliability} A possible approach for ensuring reliability is to have redundancy in terms of distributed hardware and perform the same tests on more disjoint subsets of testbed devices.

\subsection{Challenges related to the process automation}
With the implementation of the proposed framework, it is possible to realize automated system tests. However, while the CI system and process are fully automated, the detection of errors caused by the physical environment is currently manual. This is an issue specific to wireless environments and has yet to be rigorously addressed. Additionally, the developer of the test currently needs to manually select one or more testbed devices to be used in the test.

\paragraph{Automated tests} A possible approach for enabling fully automated tests is to design and develop a mechanism that would analyze spectrum sensing data in correlation with the tests and repeat the test if it failed because of a temporally jammed channel. This process could be made transparent to the developer.

\paragraph{Automated resource allocation while retaining test results reproducibility} A possible approach for enabling automated resource allocation is to develop a mechanism that would automatize the resource (i.e., testbed device) assignment for individual test execution. It would be interesting to see if the concurrent tests could be somehow categorized and how this information could be utilized by an algorithm for automatic selection of testbed devices to optimize test execution and testbed utilization while keeping test results reproducible as well.

\subsection{Challenges related to the lifecycle management of the wireless product}
Continuous software development should also consider continuous delivery of the builds in production. The proposed framework is focused on developing quality software also for restricted capability devices. However, additional consideration could be given to extending the framework with continuous delivery capabilities to solve the issue of wireless devices, which are not maintained after entering the market. Continuous delivery could also solve security issues in devices on the market. Currently, every electronics producer addresses the problem of continuous delivery separately; however, it would be beneficial to find some standard method.

\section{Summary}
\label{sec:conclusions}

In this paper, we introduced continuous integration (CI) practices for wireless network development. We proposed the COINS framework, which simplifies the code development process on real wireless testbeds and enables frequent code changes. We then showed that the proposed framework can be implemented on a testbed composed of state-of-the art hardware modules suitable for 5G capillary wireless MTC experimentation. We realized a reference implementation of COINS by integrating existing open-source components and cloud services into a fully functional system supporting CI for wireless networks. We performed a qualitative comparison of the proposed framework with three related existing frameworks. The comparison uses a list of features already used in the community for comparing experiment management tools. However, to fully illustrate the advantage of the proposed framework, we had to extend the available list of features with four CI-related ones. Finally, we identified and discussed open challenges.

\section*{Acknowledgment}
This work was partly funded by the Slovenian Research Agency (Grant no. P2-0016) and the European Community under the H2020 eWINE (Grant no. 688 116), WiSHFUL (Grant no. 645 274) and Fed4FIRE+ (Grant no. 732 638) projects.

\bibliographystyle{IEEEtran}
\bibliography{IEEEabrv,testlab-testbed-infrastructure}

\section*{Biographies}

\noindent
MATEV\v Z VU\v CNIK received his B.Sc. in electrical engineering from the University Of Ljubljana, Faculty of Electrical Engineering in 2010. He is currently a PhD candidate at the Jo\v zef Stefan International Postgraduate School and works as a senior research assistant in the Department for Communication Systems at the Jo\v zef Stefan Institute. His main interests include distributed software development and Web technologies, focusing on communication protocols and IoT integration. He has actively collaborated on several EU projects.

\bigskip\noindent
TOMA\v Z \v SOLC studied electronics at the Faculty of Electrical Engineering, University of Ljubljana and received his B.Sc. in 2007. He is currently a senior research assistant at the Department of Communication Systems, Jo\v zef Stefan Institute and a Ph.D. student at the Jo\v zef Stefan International Postgraduate School. His work involves hardware design, measurements and embedded software development as well as research in the field of spectrum sensing. He participated in various national and EU projects.

\bigskip\noindent
URBAN GREGORC received a Master degree in Electrical Engineering from the University of Ljubljana, Faculty of Electrical Engineering in 2016. After graduation, he worked at the Department of Communication Systems at the Jo\v zef Stefan Institute, Ljubljana as a research assistant and in 2018 moved to a startup ComSensus. His research is focused on Low Power Wide Area Network technologies for the Internet of Things.

\bigskip\noindent
ANDREJ HROVAT received his B.Sc. in 2004. He obtained a Ph.D. degree in Electrical Engineering from the Jo\v zef Stefan International Postgraduate School, Slovenia in 2011. He is currently a researcher in the Department of Communication Systems of the Jo\v zef Stefan Institute and an assistant professor at the Jo\v zef Stefan International Postgraduate School. His research interests include radio signal propagation, channel modelling, terrestrial and satellite fixed and mobile wireless communications, radio signal measurements and emergency communications.

\bigskip\noindent
KLEMEN BREGAR received a degree in electrical engineering from the University Of Ljubljana, Faculty of Electrical Engineering in 2013. Since 2013, he has worked as a young researcher and Ph.D. candidate at the Department of Communication Systems at Jo\v zef Stefan Institute, Ljubljana. His research interests include indoor localization, ultra-wide band communication systems, biomedical wireless sensor networks and machine learning.

\bigskip\noindent
MIHA SMOLNIKAR received his BSc from the University of Ljubljana and has more than 10 years of experience in networked embedded systems and wireless communications. His main interests include wireless sensor networks, energy harvesting, sensor technologies and the optimization of the protocol stack. In the area of IoT, he focuses on the development of products and services for the smart home, smart city and smart grid segments. He actively participates in several EU projects.

\bigskip\noindent
MIHAEL MOHOR\v CI\v C is head of the Department of Communication Systems and Scientific Counselor at the Jo\v zef Stefan Institute. He is also associate professor at the Jo\v zef Stefan International Postgraduate School. His research experience include development and performance evaluation of network protocols for mobile and wireless communication systems, cognitive radio networks, wireless sensor networks, dynamic composition of communication services and wireless experimental testbeds. He has participated in many EU projects. He is a Senior Member of IEEE.

\bigskip\noindent
CAROLINA FORTUNA received her B.Sc. in 2006, Ph.D. in 2013 and was a postdoctoral research associate at IBCN, Ghent University, 2014-2015. Currently, she is research fellow at the Department of Communication Systems, Jo\v zef Stefan Institute and an assistant at the Jo\v zef Stefan International Postgraduate School. Her research is interdisciplinary, focusing on data and knowledge driven modelling of communication and sensor systems. She has participated in EU and industrial projects with Bloomberg LP and Siemens PSE.

\end{document}